# Social Media Can Suppress or Fuel Misinformation at Its Source, Depending on Civic Engagement

Gavin Wang[1], Haofei Qin[2], Xiao Tang[3]*, Lynn Wu[4]*

[1] Daniels School of Business, Purdue University; West Lafayette, 47907, USA.

[2] Scheller College of Business, Georgia Institute of Technology; Atlanta, 30308, USA.

[3] School of Public Policy and Management, Tsinghua University; Beijing, 100190, China.

[4] The Wharton School, University of Pennsylvania; Philadelphia, 19104, USA.

* Corresponding authors. Email: Xiao Tang: tangxiao@tsinghua.edu.cn; Lynn Wu: wulynn@wharton.upenn.edu.

**Abstract:** Social media is widely viewed as a channel for spreading misinformation, but its effects may begin earlier, by shaping whether misinformation is produced in the first place. We examine whether social media suppresses or amplifies the supply of misinformation and identify civic engagement as a key boundary condition. Using data on Chinese local governments' GDP reporting from 2011 to 2019, we find that social media adoption reduces GDP misreporting *only* when civic engagement is high. This reduction is driven by posts that disclose specific economic data rather than promotional content. When civic engagement is low, however, social media adoption instead increases GDP misreporting. These findings reveal a dual upstream role of social media in the misinformation process: it can deter misinformation at its source when disclosures are exposed to engaged citizens, but can amplify its supply when civic engagement is limited.

The proliferation of social media has transformed how information is produced and disseminated. Much research has focused on its role in spreading misinformation, showing how decentralized diffusion, polarization, and echo chambers can amplify falsehoods at scale (*1-11*). Yet social media may affect misinformation even before it enters the information ecosystem—by changing organizations' incentives to produce false or misleading information in the first place. Public disclosure on social media makes organizational claims visible to large, participatory audiences, potentially exposing inconsistencies to citizens and domain experts (*12-18*). Such exposure can raise the expected reputational or political costs of deception and deter misinformation at its source.

This disciplining effect, however, may depend critically on civic engagement. Disclosure creates opportunities for scrutiny but does not guarantee that scrutiny occurs. Where citizens actively participate in public affairs, social media can make government claims easier to examine, challenge, and report, increasing the expected cost of misreporting. Where civic engagement is limited, the same technology may instead provide governments with a powerful channel for promoting misleading claims with little prospect of challenge. Social media may therefore either suppress or increase the supply of misinformation, depending on the civic environment in which disclosure occurs. Evidence from crowdsourced fact-checking and citizen challenges to official statistics demonstrates the potential for participatory audiences to identify misleading claims (*19-26*), but whether such participation changes the production of misinformation remains unclear.

We examine this question using local governments in China, where incentives to manipulate official economic statistics are well documented (*27-30*). We construct a panel of all prefecture-level governments from 2011 to 2019, combining their social media adoption and activity with officially reported GDP and economic performance inferred from satellite nighttime-light data (*29, 31, 32*). The discrepancy between reported and satellite-based GDP provides a measure of misreporting. Local governments overreport GDP growth by 2.3 percentage points on average, and social media adoption is associated with a 0.49-percentage-point decline in misreporting on average. This average, however, masks sharply divergent effects across civic environments.

We measure civic engagement using citizens' formal and informal participation in local public affairs. Social media adoption reduces GDP misreporting only where civic engagement is high; where civic engagement is low, it instead increases misreporting. Descriptive evidence from Liaoning Province illustrates the proposed scrutiny mechanism: citizen reports of government data discrepancies rose sharply following the first local government's adoption of WeChat in a province where data manipulation was subsequently publicly acknowledged. Consistent with this mechanism, reductions in GDP misreporting are concentrated in social media posts that disclose specific economic data rather than promotional content. Social media activity also does not predict manipulation of pollution data, which is subject to government manipulation but not mandated for social media disclosure (*27*). We further address potential endogeneity in both social media activity and civic engagement using separate instrumental-variable strategies.

These findings reveal a dual upstream role of social media in the misinformation process. Rather than only amplifying or correcting misinformation after it has been produced, social media can alter incentives to produce misinformation at its source. Whether it disciplines or facilitates misreporting depends on civic engagement: disclosure can translate civic participation into decentralized scrutiny when citizens are engaged, but can provide an additional channel for misleading claims when they are not. This distinction shifts attention from content moderation after misinformation enters circulation to the conditions that shape its production in the first place.

**Background of Data Misreporting in Local Chinese Governments**

China's administrative system follows a multi-tier principal-agent structure (*33*) (Fig. 1), in which higher-level governments appoint and evaluate lower-level officials based on performance targets, including GDP growth and pollution control (*34, 35*). Because meeting these targets can affect career advancement, local officials have strong incentives to report favorable performance, particularly GDP growth (*36*). Until 2020, China maintained a dual-track accounting system in which national and local GDP figures were computed independently, with local governments responsible for producing their own GDP statistics (*37*). This arrangement created opportunities for strategic misreporting. In Liaoning Province, for example, the governor publicly acknowledged in 2017 that local governments had falsified fiscal data during 2011–2014 and that other economic statistics also contained inaccuracies (*38*).

**Figure 1: Multi-Tier Principal-Agent Structure of Chinese Governments**

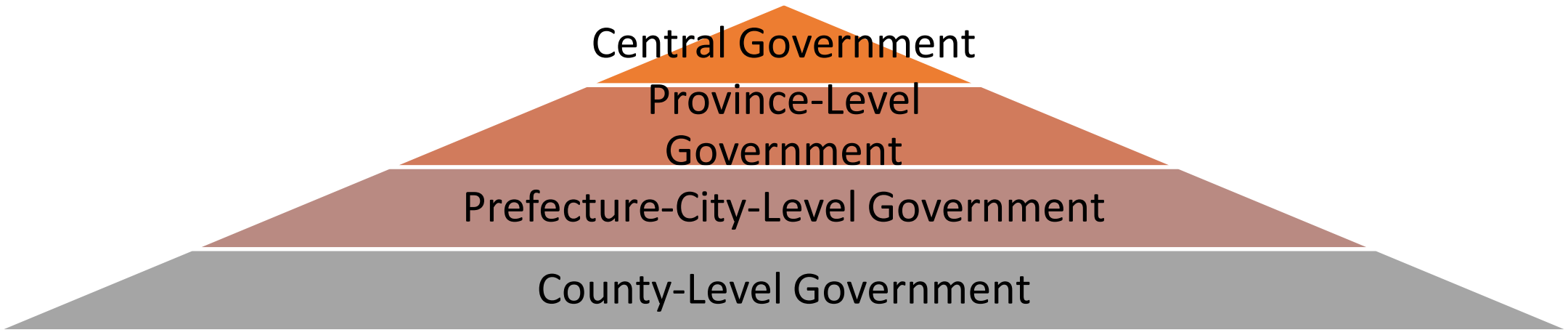

Before social media adoption, local governments primarily disseminated GDP and related statistics through newspapers and official websites, which offered limited opportunities for public engagement. Beginning in 2011, the central government required prefecture-level governments to adopt social media platforms such as WeChat for public communication, including GDP-related disclosures (*39, 40*). This mandate exposed government claims to broader, more participatory audiences and created new opportunities for citizens to examine and respond to official information. The staggered adoption of social media across local governments therefore provides a setting to examine whether greater public exposure reduces or increases the supply of misinformation—and whether its effect depends on local civic engagement.

**Data and Descriptive Analyses**

We constructed a panel dataset of 292 city-level governments in China covering the years 2011 to 2019. Our primary dependent variable is GDP manipulation, measured as the log-ratio of officially reported GDP to GDP inferred from satellite nighttime-light data (*29, 32*). We validate this measure using an alternative benchmark based on national GDP reported by the central government. Both approaches indicate that local governments overreport GDP growth by 2.3 percentage points on average. Figure 2(a) maps GDP misreporting across sample cities in 2015.

Our key explanatory variables are social media adoption, indicating whether a government has adopted WeChat for public communication, and social media activity, measured by the number of GDP-related posts on its official account. We focus on WeChat because it is the most widely adopted social media platform in China, and local governments in China are required by higher-level governments to publish information on WeChat public accounts. We further distinguish posts that disclose specific economic data from promotional posts to examine the information-disclosure mechanism. Figure 2(b) maps social media activity in 2015.

**Figure 2: Local GDP Manipulation Level (Left) and Government Social Media Activity**

**Level (Right) in Sample Cities in 2015**

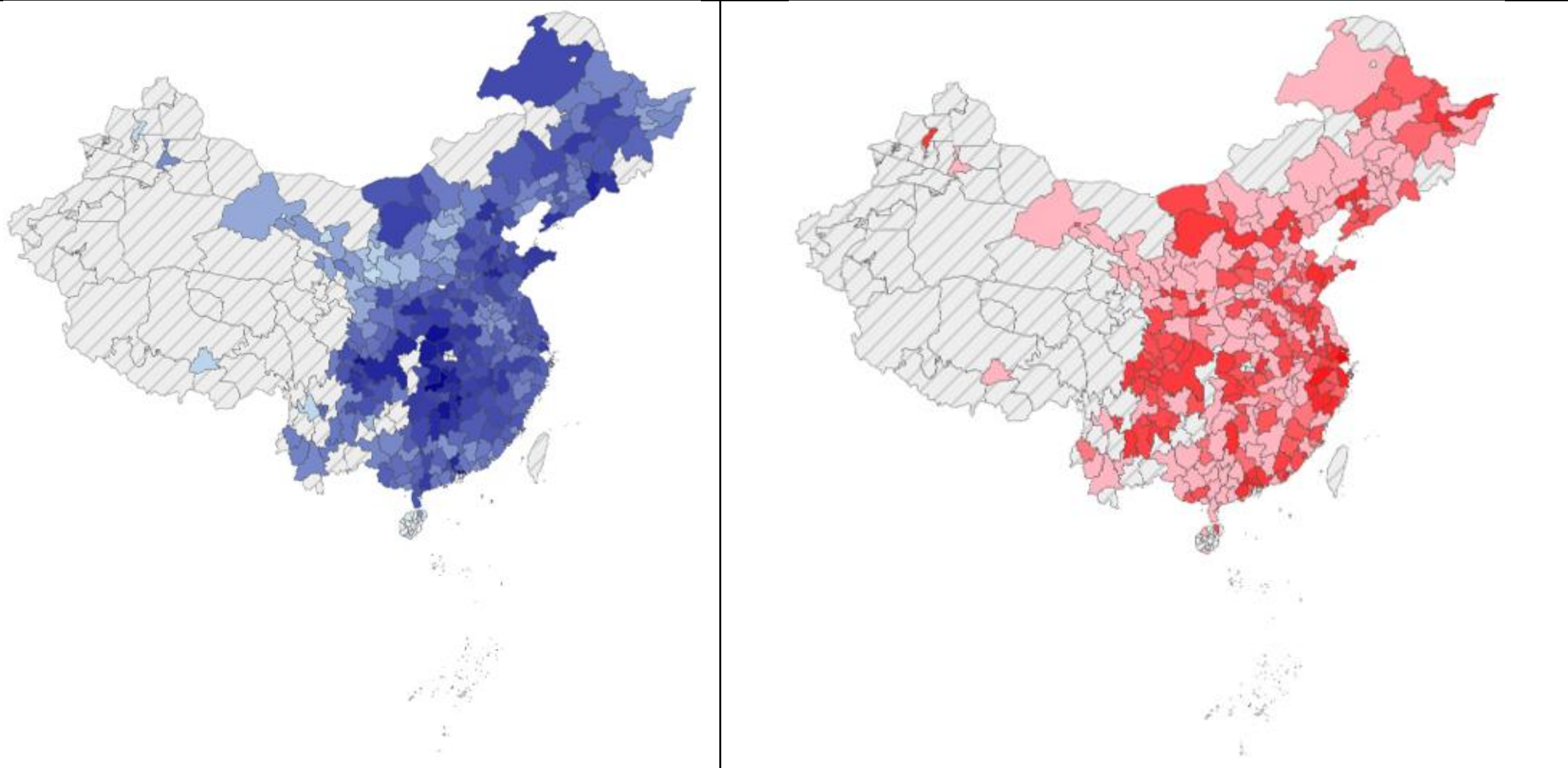

**(a) Local GDP Manipulation Level in 2015** | **(b) Government Social Media Activity Level in 2015**

Note: Colored areas represent cities in the sample, with darker shades indicating higher values. The sample covers 292 prefecture-level administrative units, including the four centrally administered municipalities (Beijing, Shanghai, Tianjin, and Chongqing). Sansha is excluded because of limited data availability. The panel is unbalanced because several prefecture-level cities, such as Linzhi, were established after 2011. Gray hatched areas indicate ethnic autonomous regions excluded from the sample because of their distinct administrative arrangements.

We measure civic engagement at the city-year level using the first principal component of two per capita measures: administrative lawsuits and citizen comments on government message boards (*41-44*). Both capture citizens' participation in public affairs through formal and informal channels. Figure 3 previews our central finding: social media adoption is associated with lower GDP manipulation where civic engagement is high but higher manipulation where civic engagement is low. Controls include population, industrial structure, economic development, retail sales, and local leadership turnover and tenure; all specifications include city and year fixed effects.

**Figure 3: Distribution of GDP Manipulation in Regions with High versus Low Civic Engagement after Social Media Adoption**

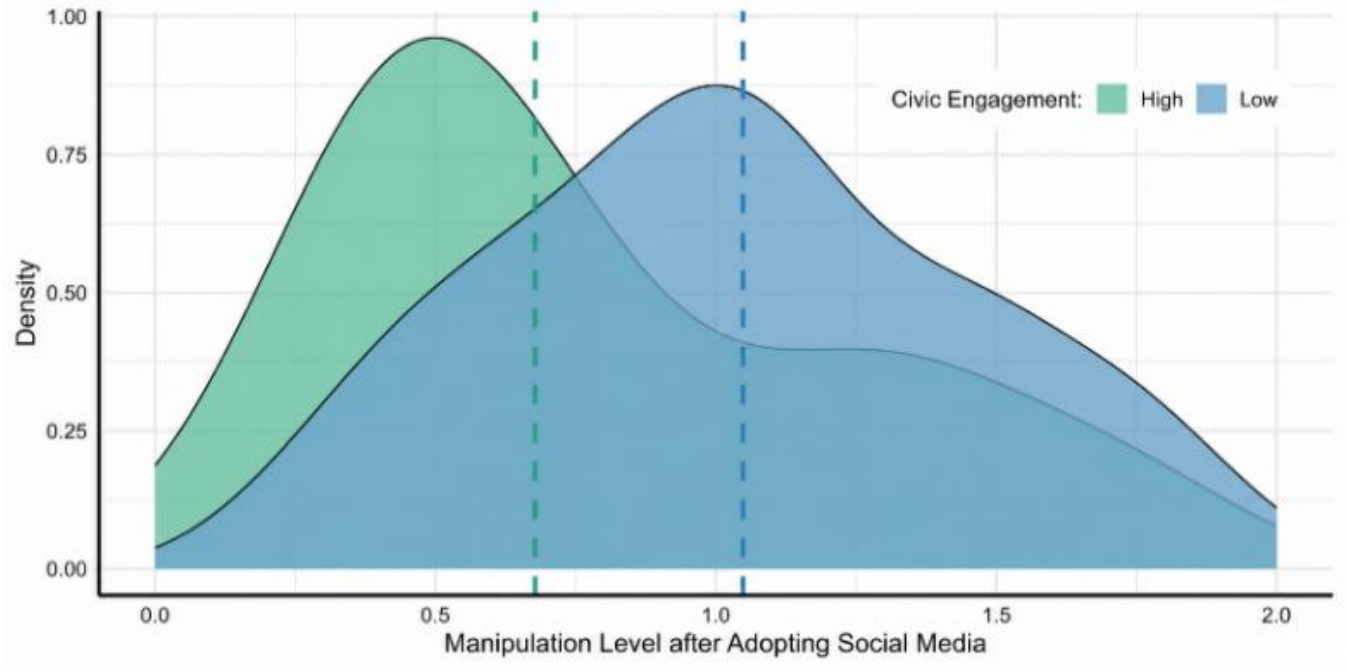

Because both social media activity and civic engagement may be endogenous, we employ separate instrumental-variable strategies. For social media activity, we use the average activity of other governments in the same province and year as a Hausman-type instrument (*45, 46*). Civic engagement varies predominantly across rather than within cities: 81.8% of its variation occurs across cities and only 18.2% within cities over time, suggesting that it largely reflects a local civic environment inherited by incumbent governments rather than one they can readily alter. Nevertheless, we further instrument civic engagement using the interaction between the annual nationwide number of investigated senior officials—a national shock that heightens public attention to political participation—and historical *Jinshi* density during the Ming–Qing period, which captures persistent cross-city differences in civic-participation traditions (*47-50*). This Bartik-style instrument exploits differential local exposure to common national political shocks based on predetermined historical differences.

## Effect of Social Media on GDP Misreporting

We first provide descriptive evidence on how government social media may facilitate citizen scrutiny of potentially misleading information. We focus on Liaoning Province, where the provincial governor publicly acknowledged in 2017 that local governments had falsified fiscal data during 2011–2014 and that other economic statistics also contained inaccuracies. Citizen reports of government data discrepancies rose sharply following the first local government's adoption of WeChat and remained above their pre-adoption level thereafter (Fig. 4). Although descriptive, this pattern is consistent with social media making government claims more visible and accessible, allowing existing civic engagement to translate into more active scrutiny of government information.

**Figure 4: Citizen Reporting of Government Data Discrepancy Following Social Media Adoption in Liaoning Province**

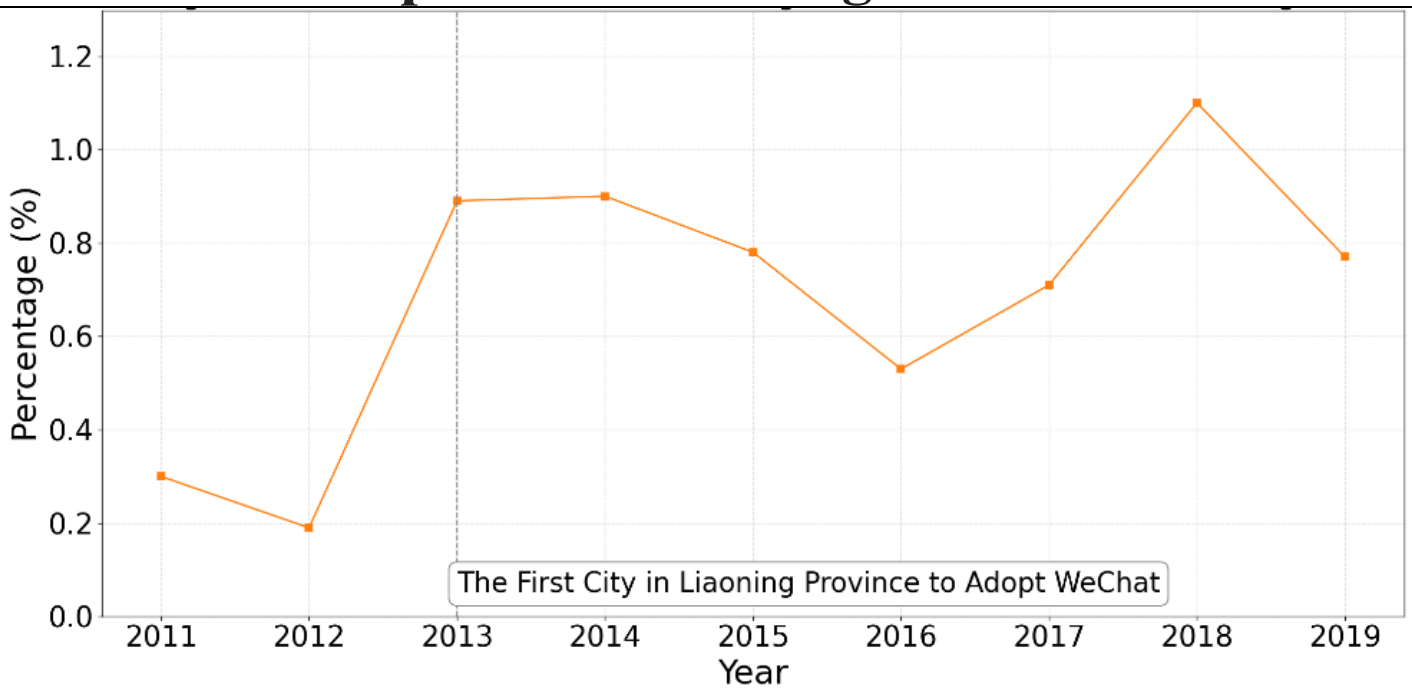

Notes: The figure plots the annual percentage of citizen comments on government message boards concerning government data discrepancies, including suspected data falsification or exaggeration. The dashed line marks 2013, when the first prefecture-level city in Liaoning adopted WeChat for official government communication.

We next examine whether this increased scope for citizen scrutiny is associated with changes in governments' supply of misinformation. Table 1 shows that, on average, social media adoption is associated with lower GDP misreporting. WeChat adoption reduces GDP inflation by 0.49 percentage point (Column 1), relative to an average overreporting of 2.3 percentage points. A 1% increase in GDP-related posts is associated with a 0.12-percentage-point reduction (Column 3). These estimates remain robust using instrumental-variable specifications (Columns 2 and 4).

The content of social media disclosure further points to a scrutiny mechanism. Posts containing specific economic data are associated with less misreporting (Columns 5 and 6), whereas promotional posts without such data are associated with more misreporting (Columns 7 and 8).

Thus, social media does not uniformly improve information quality: its disciplining effect emerges when disclosure provides citizens with verifiable information, whereas noninformational use may instead facilitate misleading performance narratives.

**Table 1: Effect of Social Media on Government Data Manipulations**

| | (1) | (2) | (3) | (4) | (5) | (6) | (7) | (8) |
|---|---|---|---|---|---|---|---|---|
| DV: GDP Manipulation | FE | FE+2SLS | FE | FE+2SLS | FE | FE+2SLS | FE | FE+2SLS |
| WeChat Adoption | -0.00486* | **-0.0303***** | | | | | | |
| | (0.00262) | **(0.00794)** | | | | | | |
| Log Number of GDP-related Articles | | | -0.00121* | **-0.00466***** | | | | |
| | | | (0.000653) | **(0.00178)** | | | | |
| Log Number of GDP-related Articles (with Data) | | | | | -0.00154** | **-0.00703***** | -0.00644*** | **-0.0373***** |
| | | | | | (0.000762) | **(0.00195)** | (0.00189) | **(0.00756)** |
| Log Number of GDP-related Articles (without Data) | | | | | | | 0.00527*** | **0.0345***** |
| | | | | | | | (0.00175) | **(0.00816)** |
| Control Variables | City FE, Year FE, Population, Population Density, GDP Per Capita, Secondary Industry Proportion, Tertiary Industry Proportion, Retail Consumption, Leadership Change, Mayor Tenure, Municipal Party Secretary Tenure | | | | | | | |
| Observations | 2,524 | 2,524 | 2,524 | 2,524 | 2,524 | 2,524 | 2,524 | 2,524 |
| R-squared | 0.660 | | 0.660 | | 0.660 | | 0.662 | |

Notes: 1. Columns (1), (3), (5), and (7) report fixed-effects estimates. The dependent variable is the GDP manipulation level. WeChat Adoption is a dummy variable that indicates whether the city has adopted WeChat to publish GDP-related information in the year. Log Number of GDP-related articles is the log number of GDP-related articles published by each government each year. GDP-related articles are further classified into articles that contain specific economic data and articles that do not contain such data.
2. Columns (2), (4), (6), and (8) use instrumental variable 2SLS. Instrumental variables are the average social media activity of other governments in the same province and year.
3. Clustered standard errors (at the city level) in parentheses: *** p<0.01, ** p<0.05, * p<0.1.

## Civic Engagement Determines Whether Social Media Suppresses or Increases Misreporting

We next examine whether the effect of social media depends on local civic engagement. The interactions between civic engagement and both WeChat adoption and social media activity are negative and statistically significant (Table 2). More importantly, the direction of social media's effect reverses across civic environments. Where civic engagement is high, social media is associated with less GDP misreporting; where civic engagement is low, it is associated with more misreporting (Fig. 5). The instrumental-variable estimates yield the same pattern (Columns 2 and 4).

Because local governments adopted WeChat at different points in time, we further examine the robustness of these findings using the staggered difference-in-differences estimator of Callaway and Sant'Anna (*51*). The dynamic estimates provide no systematic evidence of pre-treatment effects and reproduce the central heterogeneity documented above: WeChat adoption is followed by lower GDP manipulation in cities with higher civic engagement but higher manipulation in cities with lower civic engagement (fig. S2).

These results suggest that disclosure alone is insufficient to discipline misinformation. Social media can translate civic engagement into scrutiny by exposing government claims to citizens who are willing to examine and challenge them. Where such engagement is limited, however, governments may instead use social media to promote inflated economic performance with less risk of challenge. Civic engagement therefore determines whether social media constrains or facilitates the supply of misinformation at its source.

**Table 2: Civic Engagement Determines the Direction of Social Media's Effect**

| | (1) | (2) | (3) | (4) |
|---|---|---|---|---|
| DV: GDP Manipulation | FE | FE+2SLS | FE | FE+2SLS |
| Civic Engagement | 0.0125*** | 0.0572 | 0.00817*** | 0.0320 |
| | (0.00264) | (0.0351) | (0.00228) | (0.0200) |
| WeChat Adoption | -0.00473* | **-0.0208*** | | |
| | (0.00262) | **(0.0114)** | | |
| Civic Engagement * WeChat Adoption | -0.00824*** | **-0.0748*** | | |
| | (0.00191) | **(0.0451)** | | |
| Log Number of GDP-related Articles | | | -0.00208 | **-0.00213** |
| | | | (0.00139) | **(0.00576)** |
| Civic Engagement * Log Number of GDP-related Articles | | | -0.00350*** | **-0.0327*** |
| | | | (0.000793) | **(0.0169)** |
| Other Controls | City FE, Year FE, Population, Population Density, GDP Per Capita, Secondary Industry Proportion, Tertiary Industry Proportion, Retail Consumption, Leadership Change, Mayor Tenure, Municipal Party Secretary Tenure | | | |
| Observations | 2,524 | 2,126 | 2,524 | 2,126 |
| R-squared | 0.666 | | 0.665 | |

Notes: 1. Columns (1) and (3) report fixed-effects estimates. The dependent variable is the GDP manipulation level. Civic Engagement Index is demeaned and standardized. WeChat Adoption is a dummy variable that indicates whether the city has adopted WeChat to publish GDP-related information in the year. Log Number of GDP-related Articles is the log number of GDP-related articles published by each government each year.
2. Columns (2) and (4) use instrumental variable 2SLS. The instrumental variable for Civic Engagement is the interaction between the nationwide number of investigated senior officials, which varies over time, with historical Jinshi density from the Ming–Qing period, which varies across cities. The instrumental variables for social media activities are the average social media activity of other governments in the same province and year. The interaction between social media disclosure and Civic Engagement is instrumented using the product of the corresponding social-media and civic-engagement instruments.
3. Clustered standard errors (at the city level) in parentheses: *** p<0.01, ** p<0.05, * p<0.1.

**Figure 5: Social Media and GDP Misreporting Across Levels of Civic Engagement**

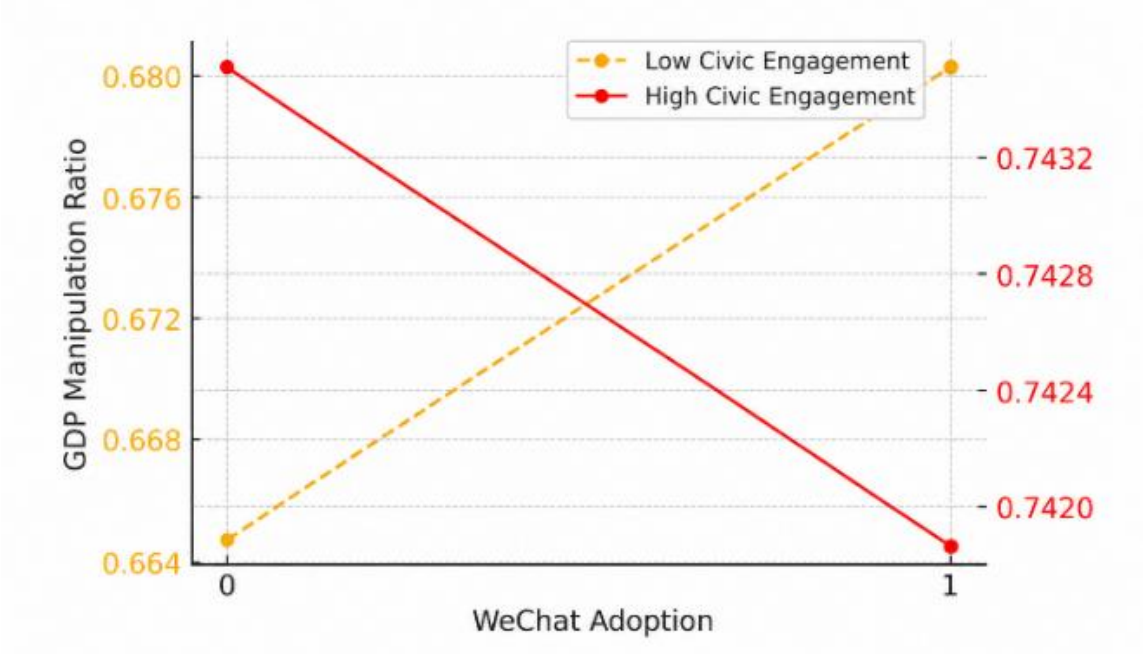

Notes: Cities are grouped by civic engagement in 2016. Solid and dashed lines show the relationship between WeChat adoption and GDP manipulation for cities above the 85th and below the 15th percentiles of civic engagement, respectively.

## Placebo Test: Pollution Emission Data Manipulation

To address the possibility that governments adopting social media are systematically less likely to manipulate data, we conduct a placebo test using pollution emissions. Chinese local governments also manipulate environmental data (*27*), but unlike GDP, these data are not mandated for social media disclosure (*52*). We measure $SO_2$ manipulation as the discrepancy between satellite-observed and officially reported emissions and test whether it is associated with government social media use in Table 3. The results show that social media adoption and activity, including emissions-related and GDP-related posts, do not significantly predict $SO_2$ manipulation under either fixed-effects or instrumental-variable specifications. This null relationship contrasts with the patterns for GDP reporting, making differences in underlying government integrity less likely to explain our main findings and supporting a mechanism tied specifically to public disclosure of economic information.

**Table 3: Placebo Test Using $SO_2$ Emissions Manipulation**

| | (1) | (2) | (3) | (4) | (5) | (6) | (7) | (8) |
|---|---|---|---|---|---|---|---|---|
| DV: $SO_2$ Manipulation | FE | FE+2SLS | FE | FE+2SLS | FE | FE+2SLS | FE | FE+2SLS |
| WeChat Adoption | -0.263 | **-0.0836** | | | | | | |
| | (0.251) | **(0.0588)** | | | | | | |
| Log Number of Emission-related Articles | | | -0.0586 | **-0.223** | | | | |
| | | | (0.0545) | **(0.220)** | | | | |
| Log Number of GDP-related Articles | | | | | -0.0512 | **-0.00212** | | |
| | | | | | (0.0457) | **(0.0240)** | | |
| Log Number of GDP-related Articles (with Data) | | | | | | | -0.0506 | **0.0199** |
| | | | | | | | (0.0446) | **(0.0412)** |
| Control Variables | City FE, Year FE, Population, Population Density, GDP Per Capita, Secondary Industry Proportion, Tertiary Industry Proportion, Retail Consumption, Leadership Change, Mayor Tenure, Municipal Party Secretary Tenure | | | | | | | |
| Observations | 2,349 | 2,349 | 2,349 | 2,349 | 2,349 | 2,349 | 2,349 | 2,349 |
| R-squared | 0.0324 | | 0.0320 | | 0.0316 | | 0.0312 | |
| First-stage F-statistic | | 146.50 | | 41.21 | | 76.23 | | 75.72 |

Notes: 1. The dependent variable is $SO_2$ emission manipulation. All specifications include city and year fixed effects and the controls listed in Table 1. The smaller sample reflects missing official $SO_2$ data for some cities.
2. Columns (2), (4), (6), and (8) report 2SLS estimates using the corresponding provincial averages as instruments for WeChat adoption, emissions-related posts, GDP-related posts, and GDP-related posts containing economic data, respectively.
3. Clustered standard errors (at the city level) in parentheses: *** $p<0.01$, ** $p<0.05$, * $p<0.1$

## Discussion

Our findings show that social media can either suppress or increase the supply of misinformation, depending on civic engagement. Among Chinese local governments, social media adoption reduces GDP misreporting where civic engagement is high but increases it where engagement is low. This divergence suggests that digital disclosure does not inherently improve information quality. Rather, social media makes government claims more visible and accessible, while civic engagement determines whether this visibility translates into scrutiny and accountability. Consistent with this mechanism, reductions in misreporting are concentrated in posts that disclose specific, verifiable economic data, and citizen reporting of data discrepancies increased following social media adoption in Liaoning.

These findings shift attention from the diffusion of misinformation to its production. Most interventions address misinformation after it has entered circulation, through fact-checking, content moderation, or corrective information. Our results point to an upstream mechanism: by changing the expected consequences of making misleading claims, social media can alter incentives to produce misinformation in the first place. Importantly, this effect is not uniformly disciplinary. When disclosures are exposed to engaged citizens, social media can facilitate decentralized scrutiny and increase the expected cost of misreporting. When civic engagement is limited, however, the same communication technology can provide an additional channel for promoting misleading claims with relatively little risk of challenge. Civic engagement thus represents a key boundary condition linking digital transparency to information integrity.

Although our evidence comes from Chinese local governments, the underlying distinction between disclosure and engagement may apply more broadly wherever organizations communicate verifiable claims directly to public audiences. Transparency technologies alone may be insufficient to improve information quality if audiences lack the incentives or capacity to scrutinize disclosed information. Future research could examine whether similar dynamics arise in corporate disclosure, nonprofit organizations, or other institutional settings, and identify the forms of civic participation that most effectively translate digital visibility into accountability. More broadly, reducing misinformation may require attention not only to how false information spreads, but also to the conditions that shape whether it is produced at all.

**Acknowledgements:** We thank seminar participants in the Workshop on Information Systems and Economics (WISE) 2023 and Statistical Challenges in E-Commerce Research (SCECR) 2024 for providing valuable feedback on early drafts if this work.

**Author Contributions:** W.G. conceived the research and collected the descriptive data. W.G. and Q.H. processed and analyzed the data with T.X. and W.L. aiding in research conceptualization and data analysis. W.G. wrote the paper with input from Q.H. T.X. and W.L. edited the paper and supervised the work. All authors made revisions and approved the final manuscript.

**Competing Interests:** The authors declare no competing interests.

**Data and Materials Availability:** Code and data to reproduce the findings in this manuscript can be accessed from https://github.com/HaofeiQin/SocialMedia-ReduceMisinformation.git.

**Supplementary Materials**

A. Materials and Methods

B. Supplementary Analyses

C. Validation and Robustness Checks

Figs. S1 to S3

Tables S1 to S8

References (S*1*–S*26*)

# Supplementary Materials for

## Social Media Can Suppress or Fuel Misinformation at Its Source, Depending on Civic Engagement

Gavin Wang[1], Haofei Qin[2], Xiao Tang[3]*, Lynn Wu[4]*

[1] Daniels School of Management, Purdue University; West Lafayette, 47907, USA.
[2] Scheller College of Business, Georgia Institute of Technology; Atlanta, 30308, USA.
[3] School of Public Policy and Management, Tsinghua University; Beijing, 100190, China.
[4] The Wharton School, University of Pennsylvania; Philadelphia, 19104, USA.

* Corresponding authors. Email: Xiao Tang: tangxiao@tsinghua.edu.cn; Lynn Wu: wulynn@wharton.upenn.edu.

**The PDF file includes:**

## A. Materials and Methods

This section describes the construction of the key variables and the empirical strategies used in our analyses. We first describe the measurement of local governments' GDP manipulation and social media disclosure, followed by the construction of our measure of civic engagement and other control variables. We then present the empirical specifications used to estimate the relationship between social media disclosure and GDP manipulation, as well as the moderating role of civic engagement. Finally, we discuss our identification strategies for addressing potential endogeneity concerns. Additional analyses addressing staggered social media adoption, heterogeneity across levels of civic engagement, measurement validation, and robustness are reported in subsequent sections of the Supplementary Materials.

### A1. Measuring Local Government GDP Manipulation

To measure GDP manipulation by local governments, we combine officially reported GDP figures with satellite-based nighttime light intensity as an independent proxy for local economic activity. Official GDP data are collected from statistical yearbooks published by municipal governments, which report annual economic output and other socioeconomic indicators. Additional details on the collection of reported GDP data are provided in Section C4.

We use nighttime light intensity as an external benchmark for underlying economic activity. A large literature documents a strong relationship between nighttime luminosity and economic output (*1-3*). Although nighttime light data are an imperfect proxy for economic activity, advances in satellite imagery and data-processing techniques have substantially improved their coverage and reliability (*4*).

To further improve measurement precision, we use a recently developed nighttime light dataset constructed using a convolutional long short-term memory (LSTM) approach (*5*). The dataset provides annual nighttime light intensity at a 1 km × 1 km resolution across China. We aggregate pixel-level observations within prefecture-level city boundaries to construct annual city-level nighttime light intensity.

Following prior research (*3, 6*), we construct our primary measure of GDP manipulation as:

$$\textit{GDP Manipulation}_{i,t} = \log(\textit{Reported GDP}_{i,t}) / \log(\textit{Nighttime Light}_{i,t}).$$

The measure captures the extent to which officially reported GDP is high relative to economic activity reflected in nighttime luminosity. Higher values indicate greater potential over-reporting of local GDP. Because the measure relies on the relationship between reported economic activity and an independently observed proxy, it does not require us to observe the true level of city GDP (*3, 6*).

We conduct several additional analyses to assess the validity and robustness of this measure. These include an alternative rank-based measure of GDP manipulation, comparisons with officially documented cases of GDP falsification (*7*), and an alternative estimate of local GDP using national GDP and nighttime light intensity as benchmarks. These analyses are reported in Section C.

### A2. Measuring Social Media Disclosure

WeChat is the most widely adopted social media platform in China (*8*), and local governments are required by higher-level authorities to disseminate government information, including GDP-related information, through their official WeChat accounts (*9, 10*). To measure local governments' social media disclosure, we collect the full corpus of articles published by their official WeChat accounts, including the full text and publication date of each article.

We construct two types of measures to capture local governments' use of social media. First, WeChat Adoption is an indicator equal to one if a city government has adopted an official WeChat account for public communication in a given year, and zero otherwise. Second, to capture the intensity of GDP-related disclosure, we count the number of GDP-related articles published by each local government annually. We identify GDP-related articles using a comprehensive keyword dictionary reported in Table S1. Because the distribution of article counts is highly skewed, we use the log-transformed number of GDP-related articles, ***Log Number of GDP-related Articles***, in our empirical analyses.

Importantly, not all GDP-related articles provide substantive economic information. Keyword-based identification may capture both articles that disclose specific economic data and articles that discuss economic development without reporting verifiable quantitative information. Figure S1 provides examples of these two types of articles. We therefore manually review the identified GDP-related articles and classify them into two categories: ***GDP-related Articles (with Data)***, which contain specific economic data, and ***GDP-related Articles (without Data)***, which do not contain such data. We then construct the corresponding log-transformed annual counts, ***Log Number of GDP-related Articles (with Data)*** and ***Log Number of GDP-related Articles (without Data)***.

This distinction allows us to examine whether the relationship between social media disclosure and GDP manipulation is associated specifically with the disclosure of verifiable economic information, rather than with government social media activity more generally. If social media reduces misinformation by increasing the visibility and verifiability of government-reported information, we would expect the effect to be concentrated among GDP-related articles that contain specific economic data.

### A3. Measuring Civic Engagement

To measure the level of civic engagement in each city, we construct a composite index based on two indicators of citizens' participation in public affairs and interactions with local governments: (1) the annual number of administrative litigation cases per capita and (2) the annual number of government message-board comments per capita.

Administrative litigation refers to legal proceedings in which citizens or organizations challenge administrative decisions made by government agencies. The use of this formal institutional channel reflects citizens' willingness to actively engage with government actions and defend their interests through legal means (*11, 12*). Government message-board comments capture a complementary form of civic engagement through less formal and more accessible channels. These comments are citizen-generated posts on official online platforms hosted by local governments, through which residents raise concerns, request information, provide feedback, and communicate with public officials about government policies and public services (*13, 14*). Together, the two measures capture citizens' propensity to actively engage with local governments through both formal and informal channels.

Because the two indicators capture related dimensions of civic engagement, we combine them using principal component analysis (PCA). Specifically, we standardize the two measures and use the first principal component to construct the ***Civic Engagement Index***. The index is standardized to have a mean of zero and a standard deviation of one, with higher values indicating greater civic engagement.

Civic engagement is theoretically relevant to our setting because social media disclosure does not automatically translate into effective public oversight. Although citizens generally do not possess the information or technical expertise required to independently reconstruct local GDP statistics, engaged citizens are more likely to pay attention to government-reported information, compare official claims with locally observable economic conditions, raise questions when inconsistencies become salient, and communicate concerns through available institutional channels (*15, 16*). Civic engagement therefore increases the likelihood that information disclosed through social media is subject to public attention and potential scrutiny.

This mechanism is consistent with evidence that citizen participation can contribute to the detection of statistical misconduct. For example, Yuqin Liu, Director of the Law Enforcement and Supervision Bureau at the National Bureau of Statistics of China, noted that citizen reports have constituted an important source of leads regarding statistical falsification (*17*). We therefore view civic engagement not as a direct measure of citizens' ability to verify GDP statistics, but as a measure of the local social environment in which government disclosures are more likely to attract attention, questioning, and feedback.

We further examine the extent to which the Civic Engagement Index captures persistent differences across local civic environments rather than primarily short-term fluctuations within cities. Table S3 decomposes the variation in the index into between-city and within-city components. Between-city differences account for approximately 81.8% of the total variation, while within-city changes account for approximately 18.2%. The null hypothesis of equal mean civic engagement across cities is strongly rejected ($F = 35.39$, $p < 0.001$). These results indicate substantial and persistent heterogeneity in civic engagement across cities, while the remaining within-city variation provides meaningful temporal variation for our panel analyses.

### A4. Control Variables

To account for time-varying differences in local socioeconomic conditions that may be associated with both social media disclosure and GDP reporting, we include a set of city-level control variables. Demographic controls include ***Population*** and ***Population Density***. Economic controls include ***GDP Per Capita***, ***Secondary Industry Proportion***, ***Tertiary Industry Proportion***, and ***Retail Consumption***.

We additionally control for changes in local political leadership, which may affect both government communication practices and incentives surrounding economic reporting. We compile annual information on each city's mayor and municipal Party secretary. The municipal Party secretary is generally the highest-ranking local official, while the mayor is responsible for major administrative functions and policy implementation. We construct ***Leadership Change***, an indicator equal to one if either the mayor or municipal Party secretary changes in a given year, and zero otherwise. We also control for ***Mayor Tenure*** and ***Municipal Party Secretary Tenure***, measured as the number of years each official has held the position.

Summary statistics and definitions of the variables used in our analyses are reported in Table S2.

### A5. Empirical Specifications

We estimate the relationship between local governments' social media disclosure and GDP manipulation using the following two-way fixed-effects specification:

$$GDPManipulation_{i,t} = \beta_1 SocialMediaDisclosure_{i,t} + X_{i,t}'\theta + \tau_t + \gamma_i + \varepsilon_{i,t}$$

where $i$ indexes cities and $t$ indexes years. $GDPManipulation_{i,t}$ is the GDP manipulation measure described in Section A1, and $SocialMediaDisclosure_{i,t}$ represents the alternative measures of social media disclosure described in Section A2, including WeChat adoption and the volume of GDP-related articles. $X_{i,t}$ denotes the vector of time-varying city-level controls described in Section A4. $\gamma_i$ and $\tau_t$ represent city and year fixed effects, respectively. Standard errors are clustered at the city level.

The city fixed effects absorb time-invariant differences across cities, while the year fixed effects account for common shocks affecting all cities in a given year. The coefficient $\beta_1$ therefore captures the within-city association between changes in social media disclosure and changes in GDP manipulation, conditional on the included controls and fixed effects. Because higher values of our dependent variable indicate greater GDP manipulation, a negative $\beta_1$ indicates that greater social media disclosure is associated with lower GDP manipulation.

The timing of our measures also reflects the sequence of disclosure and GDP reporting in our setting. Social media activity indexed to year $t$ occurs during that year, whereas official GDP for year $t$ is finalized and publicly reported after the corresponding economic activity has occurred. Thus, the social media disclosure measure captures government communication occurring during the period over which the corresponding GDP figure is generated and subsequently reported. Nevertheless, Equation (S1) alone does not eliminate potential concerns about endogenous social media activity or time-varying unobserved factors. We address these concerns using the identification strategies described in Section A6 and the staggered-adoption analysis described in Section B1.

To examine whether the relationship between social media disclosure and GDP manipulation varies with local civic engagement, we estimate the following interaction specification:

$$GDPManipulation_{i,t} = \beta_1 SocialMediaDisclosure_{i,t} + \beta_2 CivicEngagement_{i,t} + \beta_3 SocialMediaDisclosure_{i,t} \times CivicEngagement_{i,t} + X_{i,t}'\theta + \tau_t + \gamma_i + \varepsilon_{i,t}$$

$CivicEngagement_{i,t}$ is the standardized Civic Engagement Index described in Section A3. The coefficient $\beta_3$ captures how the relationship between social media disclosure and GDP manipulation varies with the level of civic engagement. A negative $\beta_3$ indicates that the association between social media disclosure and GDP manipulation becomes more negative as civic engagement increases. We further examine this heterogeneity by estimating marginal effects across different levels of civic engagement and by conducting subgroup analyses, as reported in Section B2.

### A6. Identification Strategies

Although official social media adoption was promoted by higher-level government directives (*9, 10*), local governments may retain discretion over the timing and intensity of their social media activities. This raises potential concerns that time-varying local characteristics may jointly affect social media disclosure and GDP reporting behavior. We employ several complementary strategies to address these identification concerns.

First, our baseline specifications include city and year fixed effects together with a rich set of time-varying socioeconomic and political controls. City fixed effects absorb persistent differences across local governments, while year fixed effects account for common national shocks. The time-varying controls further account for changes in local demographic and economic conditions and political leadership that may be correlated with both government communication and GDP reporting. These controls do not eliminate all potential sources of endogeneity, motivating the additional analyses described below.

***Placebo outcome.*** We examine whether government social media activity predicts manipulation in a domain for which the corresponding information was not subject to the same social-media disclosure requirement. Prior research documents manipulation of local pollution-emission statistics (*6*), but higher-level government directives did not similarly require local governments to disclose pollution-emission information through official WeChat accounts (*10*). If our GDP results primarily reflect an underlying tendency for more transparent or better-governed local governments both to use social media more actively and to manipulate official statistics less, we might expect a similar relationship between social media activity and manipulation of pollution-emission data.

We construct a measure of sulfur dioxide ($SO_2$) emission manipulation by combining officially reported local $SO_2$ emissions with satellite-based measures of $SO_2$ emissions. Satellite observations are obtained from the European Space Agency's Global Emission project and measure $SO_2$ emissions per unit area at a 25 km × 25 km resolution. We aggregate these observations to the city level and compare them with officially reported $SO_2$ emissions collected from local statistical yearbooks. Because local governments have incentives to under-report pollution emissions rather than over-report them, we construct the pollution manipulation measure such that higher values indicate greater potential manipulation. We also identify pollution-related WeChat articles using the keyword dictionary reported in Table S1. The placebo estimates are reported in Table 3 of the main text. The absence of a comparable relationship in this placebo setting is inconsistent with the alternative explanation that inherently more transparent or less manipulative governments are simply more active on social media.

***Instrumental-variable strategy for social media disclosure.*** As an additional approach to addressing endogenous social media activity, we construct a leave-one-out provincial peer measure of social media activity. For each city-year, the instrument is the average social media activity of other cities in the same province and year, excluding the focal city. Local governments within the same province operate under common provincial administrative structures and are exposed to similar policy guidance and technology-adoption environments, making the social media activities of provincial peers predictive of the focal city's social media activity (*18-21*).

The identifying assumption is that, conditional on city and year fixed effects and the included time-varying controls, social media activity among other cities in the same province affects the focal city's GDP manipulation primarily through its association with the focal city's own social media activity. We assess instrument relevance using the corresponding first-stage estimates and F-statistics reported with the 2SLS results.

***Instrumental-variable strategy for civic engagement.*** We additionally construct a Bartik-style instrument to address potential endogeneity in local civic engagement. The instrument interacts time-series variation in the nationwide number of investigated senior officials with cross-sectional variation in historical *Jinshi* density during the Ming–Qing imperial examination period (*22-25*), scaled by the logarithm of contemporaneous city population. Historical *Jinshi* density captures cross-sectional variation in historical political participation associated with the imperial examination system, whereas nationwide anti-corruption investigations provide a common time-varying shock that may differentially activate contemporary civic engagement across cities. The interaction therefore generates differential temporal variation in civic engagement based on predetermined local historical characteristics.

In specifications examining the moderating role of civic engagement, both social media disclosure and civic engagement may be endogenous, and their interaction consequently requires an instrument as well. We therefore instrument social media disclosure using the leave-one-out provincial peer measure described above and civic engagement using the Bartik-style instrument. We instrument the interaction between social media disclosure and civic engagement using the product of these two instruments. Thus, the 2SLS specification jointly instruments the two constituent variables and their interaction. The identifying assumptions are that, conditional on city and year fixed effects and the included time-varying controls, the two instruments affect local GDP manipulation primarily through their respective endogenous regressors and that their product provides exogenous variation in the interaction between social media disclosure and civic engagement. We assess instrument relevance using the corresponding first-stage F-statistics reported with the 2SLS results.

Finally, because official WeChat adoption occurred at different points in time across cities, conventional two-way fixed-effects estimates may be sensitive to heterogeneous treatment effects under staggered treatment timing. We address this concern separately using the staggered difference-in-differences estimator of Callaway and Sant'Anna (*26*), as described in Section B1.

## B. Supplementary Analyses

This section presents additional analyses that further examine the relationship between social media disclosure and GDP manipulation. We first account for the staggered timing of official WeChat adoption using the Callaway and Sant'Anna difference-in-differences estimator and examine the dynamic effects of adoption. We then explore heterogeneity in the effect of social media disclosure across different levels of civic engagement. Together, these analyses provide additional evidence on both the temporal dynamics and the heterogeneous effects of social media disclosure.

### B1. Staggered Adoption and Dynamic Treatment Effects

Our baseline specifications include city and year fixed effects to exploit within-city changes in social media disclosure over time. However, local governments adopted official WeChat accounts at different points in time. Under staggered treatment adoption, conventional two-way fixed-effects estimators may be sensitive to heterogeneous treatment effects across cohorts and over time. We therefore complement our baseline analyses using the staggered difference-in-differences estimator developed by Callaway and Sant'Anna (*26*), which explicitly accommodates variation in treatment timing and treatment-effect heterogeneity.

We define treatment as a city government's adoption of an official WeChat account and estimate group-time average treatment effects based on the timing of initial adoption, using not-yet-treated cities as the comparison group. We then aggregate these estimates by event time to examine the dynamic effects of WeChat adoption before and after treatment. Figure S2, Panel A reports the event-study estimates for the full sample. Panels B and C repeat the analysis separately for cities with high and low levels of civic engagement, defined based on the sample median of the Civic Engagement Index, respectively.

The results are consistent with our baseline findings. In the full sample, the estimated effect of WeChat adoption on GDP manipulation becomes more negative following adoption. The event-study estimates also provide no systematic evidence of pre-treatment effects. More importantly, the subgroup analyses reveal substantial heterogeneity by civic engagement. Among cities with higher civic engagement, WeChat adoption is followed by a reduction in GDP manipulation. In contrast, among cities with lower civic engagement, the estimated post-adoption effects are reversed, with WeChat adoption associated with an increase in GDP manipulation. These contrasting patterns are consistent with our argument that the consequences of social media disclosure depend critically on the local civic environment: social media disclosure is associated with less GDP manipulation where civic engagement is relatively high, but with greater manipulation where civic engagement is relatively low.

### B2. Heterogeneity across Civic Engagement Levels

To further examine how the relationship between social media disclosure and GDP manipulation varies with the local civic environment, we divide observations into four groups based on the quartiles of the Civic Engagement Index and estimate the effect of social media disclosure separately within each quartile. This analysis allows us to examine whether the contrasting effects documented above reflect a broader and systematic pattern of heterogeneity rather than being driven by a particular median-based sample split.

Figure S3 reports the estimated effects across the four civic engagement quartiles. The results reveal a clear pattern: the estimated effect of social media disclosure becomes progressively more negative as civic engagement increases. At relatively low levels of civic engagement, greater social media disclosure is associated with increased GDP manipulation. As civic engagement increases, this relationship weakens and eventually reverses, such that greater social media disclosure is associated with reduced GDP manipulation at higher levels of civic engagement.

This pattern is consistent with both the interaction estimates reported in the main text and the staggered difference-in-differences results reported in Section B1. Importantly, civic engagement does not merely strengthen an otherwise uniformly negative relationship between social media disclosure and GDP manipulation. Rather, the direction of the relationship changes across civic environments: social media disclosure is associated with greater manipulation in relatively low-engagement environments but with less manipulation in relatively high-engagement environments.

## C. Validation and Robustness Checks

This section presents a series of analyses assessing the validity of our GDP manipulation measure and the robustness of our main findings. We first examine an alternative rank-based measure of GDP manipulation and assess whether our primary measure behaves as expected in documented cases of local GDP falsification. We then construct an alternative benchmark for local economic output using national GDP and nighttime light intensity. Finally, we address a temporary change in the format of local GDP reporting in 2017 by re-estimating our main analyses after excluding observations from that year.

### C1. Alternative Measure of GDP Manipulation

Our primary measure of GDP manipulation compares officially reported GDP with nighttime light intensity. To assess whether our findings depend on the specific functional form of this measure, we construct an alternative rank-based measure that relies only on the relative positions of cities in the distributions of reported GDP and nighttime light intensity.

Specifically, for each year, we separately rank cities according to their reported GDP and nighttime light intensity and construct the following measure:

$$GDPRankingDifference_{i,t} = Rank(Nighttime\ Light_{i,t}) - Rank(Reported\ GDP_{i,t})$$

The ranking is constructed such that cities with higher reported GDP or nighttime light intensity receive lower numerical ranks, with rank 1 representing the highest value. The measure captures discrepancies between a city's relative position based on officially reported economic output and its relative position based on independently observed nighttime luminosity. Positive values indicate that a city ranks relatively higher in reported GDP than would be suggested by its nighttime light intensity and therefore indicate greater potential GDP over-reporting.

This rank-based measure has the advantage of imposing fewer assumptions on the functional relationship between nighttime light intensity and economic output. In particular, it does not require the relationship between the two measures to be identical across cities or to follow the functional form embedded in our primary measure.

We re-estimate our main specifications using this alternative measure as the dependent variable. The results, reported in Table S4, are consistent with those obtained using our primary measure of GDP manipulation. Social media disclosure, particularly the disclosure of GDP-related articles containing specific economic data, is associated with lower GDP manipulation, whereas articles without such data do not exhibit the same pattern. These results indicate that our main findings are not driven by the particular construction of the baseline GDP manipulation measure.

### C2. Validation Using Documented GDP Falsification Cases

To further assess whether our GDP manipulation measure captures actual instances of statistical misreporting, we examine its behavior around documented cases of local GDP falsification. If the measure is informative, it should increase during periods in which local officials are known to have engaged in GDP manipulation.

We focus on two documented cases. The first involves Min Wang, who served as Party secretary of Liaoning Province from 2009 to 2015. According to official disciplinary records, Wang was subsequently investigated and found to have been involved in the falsification of economic statistics during his tenure. Consistent with this documented case, our GDP manipulation measure increases in Liaoning after 2010, with an average increase of approximately 2.46 percentage points relative to the preceding period.

The second case involves Jinghua Zhang, who served as Party secretary of Zhenjiang from 2012 to 2017 and was subsequently investigated for misconduct that included the falsification of economic statistics. Our measure similarly increases in Zhenjiang beginning in 2012, with an average increase of approximately 1.15 percentage points relative to the preceding period.

These case-based comparisons provide complementary validation for our measure. In both settings, the GDP manipulation measure increases during periods associated with independently documented statistical falsification. Although these cases do not provide a comprehensive validation of the measure across all cities and years, they indicate that the measure moves in the expected direction in settings where GDP falsification has been independently documented.

### C3. Using National GDP as an External Benchmark

As an additional validation exercise, we construct an alternative benchmark for local economic output using China's national GDP and nighttime light intensity. This approach is motivated by the view that national GDP statistics are generally subject to different reporting and aggregation processes than prefecture-level GDP statistics and can therefore provide an alternative benchmark for evaluating local GDP reporting.

Specifically, we estimate the relationship between national GDP and nighttime light intensity and use this relationship to predict economic output at the local level. We consider alternative model specifications and evaluate their predictive performance using mean absolute error (MAE). The best-performing specification achieves an MAE of approximately 4.03%, indicating that nighttime light intensity provides substantial predictive information about aggregate economic activity.

We then apply the estimated relationship to prefecture-level nighttime light intensity to construct predicted local GDP. Comparing predicted GDP with officially reported local GDP provides an alternative estimate of local GDP misreporting. Using this approach, we find that officially reported local GDP exceeds predicted GDP by approximately 2.3 percentage points on average between 2011 and 2019. This pattern is consistent with systematic upward bias in local GDP reporting and provides additional support for the validity of our primary GDP manipulation measure.

This exercise should nevertheless be interpreted as a complementary benchmark rather than a direct estimate of true local GDP. National GDP itself may contain measurement error, and the relationship between nighttime light intensity and economic activity estimated at the national level may not map perfectly onto prefecture-level economies. Moreover, to the extent that national GDP is itself upward biased, the estimated discrepancy between reported and predicted local GDP would tend to understate rather than overstate the extent of local over-reporting. We therefore use this analysis primarily to assess whether an independently constructed benchmark produces a pattern consistent with our primary measure.

### C4. Robustness to the 2017 Reporting-Format Change

During our data collection, we identified a temporary change in the format of local GDP reporting in 2017. In that year, statistical yearbooks for nearly all cities in our sample shifted from reporting aggregate GDP directly at the city level to reporting GDP separately at the district level. To construct city-level GDP for 2017, we therefore aggregate the reported GDP figures across districts within each city. The reporting format returned to city-level aggregates in 2018. Although this aggregation procedure allows us to construct a comparable city-level measure for 2017, the temporary change in reporting format may introduce additional measurement error or inconsistencies relative to other years.

Our baseline analyses retain the 2017 observations because the reporting-format change affected cities broadly and our specifications include year fixed effects, which absorb common shocks specific to 2017. Nevertheless, year fixed effects may not fully address measurement error arising from differences in the aggregation and reporting of district-level statistics across cities. We therefore re-estimate our main analyses after excluding all observations from 2017.

The results remain consistent with our main findings. Excluding 2017 does not materially alter the estimated relationship between social media disclosure and GDP manipulation or its heterogeneity by civic engagement. The pollution-emission placebo test also yields conclusions consistent with the baseline analysis, and the results remain robust when we combine the exclusion of 2017 with the alternative rank-based measure of GDP manipulation. Taken together, these analyses indicate that our findings are unlikely to be driven by the temporary change in GDP reporting format in 2017.

| Figure S1: Examples of GDP-related WeChat Articles posted by Local Governments | | |
|---|---|---|
| **Type of Article** | **Article Screenshot** | **Translation in English** |
| GDP-related Articles (with Data) | 郑州前三季度GDP成绩公布→<br>郑州发布 2023年10月31日 23:10 河南<br>40人听过<br>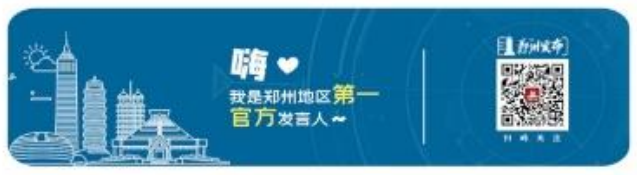<br>郑州前三季度<br>全市地区生产总值<br>完成10435.9亿元<br>同比增长6.5%<br>↓↓↓<br>郑州市统计局今日发布前三季度经济运行分析：<br>根据地区生产总值统一核算结果，前三季度，全市地区生产总值完成10435.9亿元，按不变价格计算，同比增长6.5%。<br>其中，第一产业增加值153.7亿元，同比增长0.4%；第二产业增加值4212.2亿元，增长9.1%；第三产业增加值6070.0亿元，增长4.9%。 | Zhengzhou first three quarters GDP results announced →<br>Posted by Zhengzhou at 23:10 Henan on Oct 31, 2023<br>40 heard<br>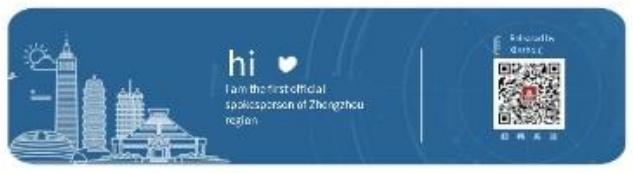<br>In the first three quarters of Zhengzhou, the city's gross regional product reached 1,043.59 billion yuan<br>Year-on-year growth of 6.5 percent<br>↓↓↓<br>Zhengzhou Statistics Bureau today released an analysis of economic operations in the first three quarters:<br>According to the unified accounting results of the regional GDP, in the first three quarters, the city's regional GDP completed 1,043.59 billion yuan, calculated at constant prices, a year-on-year increase 6.5 percent.<br>Of this total, the value added of the primary industry was 15.37 billion yuan, up 0.4 percent year-on-year; That of the secondary industry was 421.22 billion yuan, up by 9.1 percent; And that of the tertiary industry was 607.0.0 billion yuan, up by 4.9 percent. |
| GDP-related Articles (without Data) | 全市重大项目观摩暨前三季度经济运行分析会召开<br>郑州发布 2024年10月28日 10:35 河南<br>28人听过<br>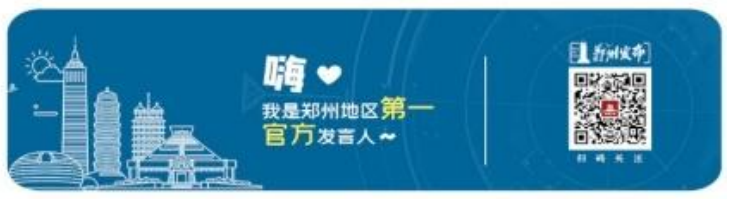<br>**全市重大项目观摩暨前三季度经济运行分析会召开**<br>**锚定目标对标对表 实干苦干全力以赴**<br>**确保经济社会发展实现"全年红"**<br>10月28日，全市重大项目观摩暨前三季度经济运行分析会召开，深入学习贯彻党的二十届三中全会和中央政治局会议精神，落实省委各省辖市工作汇报会要求，安排部署四季度经济工作。省委常委、市委书记安伟出席并讲话。 | The city's major project observation and the first three quarters of economic operation analysis will be held<br>Zhengzhou released 2024 108288 10:35 Henan<br>28 have heard<br>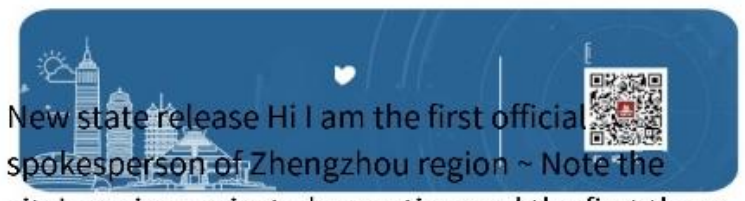<br>New state release Hi I am the first official spokesperson of Zhengzhou region ~ Note the city's major project observation and the first three quarters of economic operation analysis will be held to anchor the target against the table, hard work and go all out to ensure that economic and social development to achieve "year-round red"<br>On October 28, the city's major project observation and the first three quarters of economic operation analysis will be held, in-depth study and implementation of the Third plenary session of the 20th Central Committee of the Party and the spirit of the Central Political Bureau meeting, the implementation of the provincial committee of the provincial cities work report requirements, arrangements for the deployment of four quarters of economic work. Provincial Party Committee, Party secretary An Wei attended and made a speech. |

**Figure S2: Dynamic Effects of Official WeChat Adoption on GDP Manipulation**

**Panel A: Full Sample**

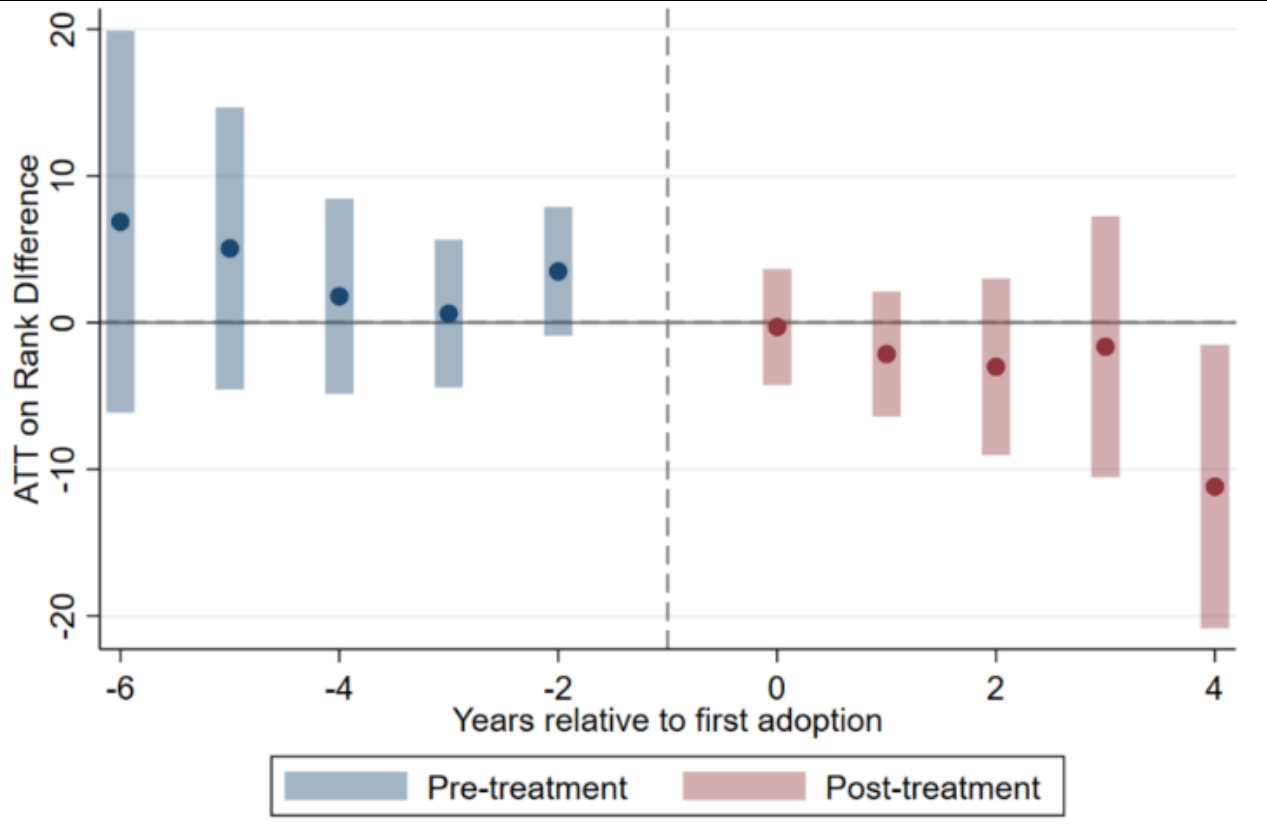

**Panel B: Sample Cities with High Civic Engagement**

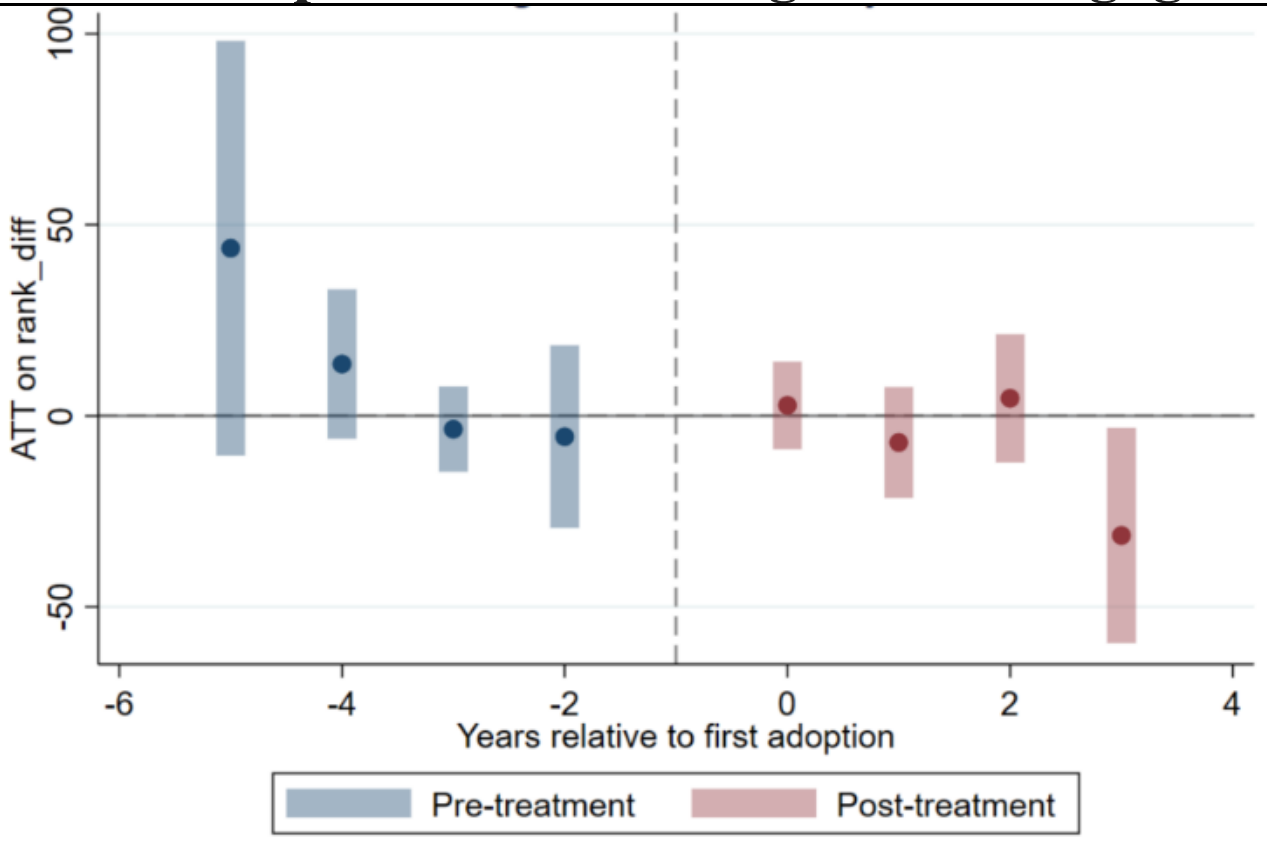

**Panel C: Sample Cities with Low Civic Engagement**

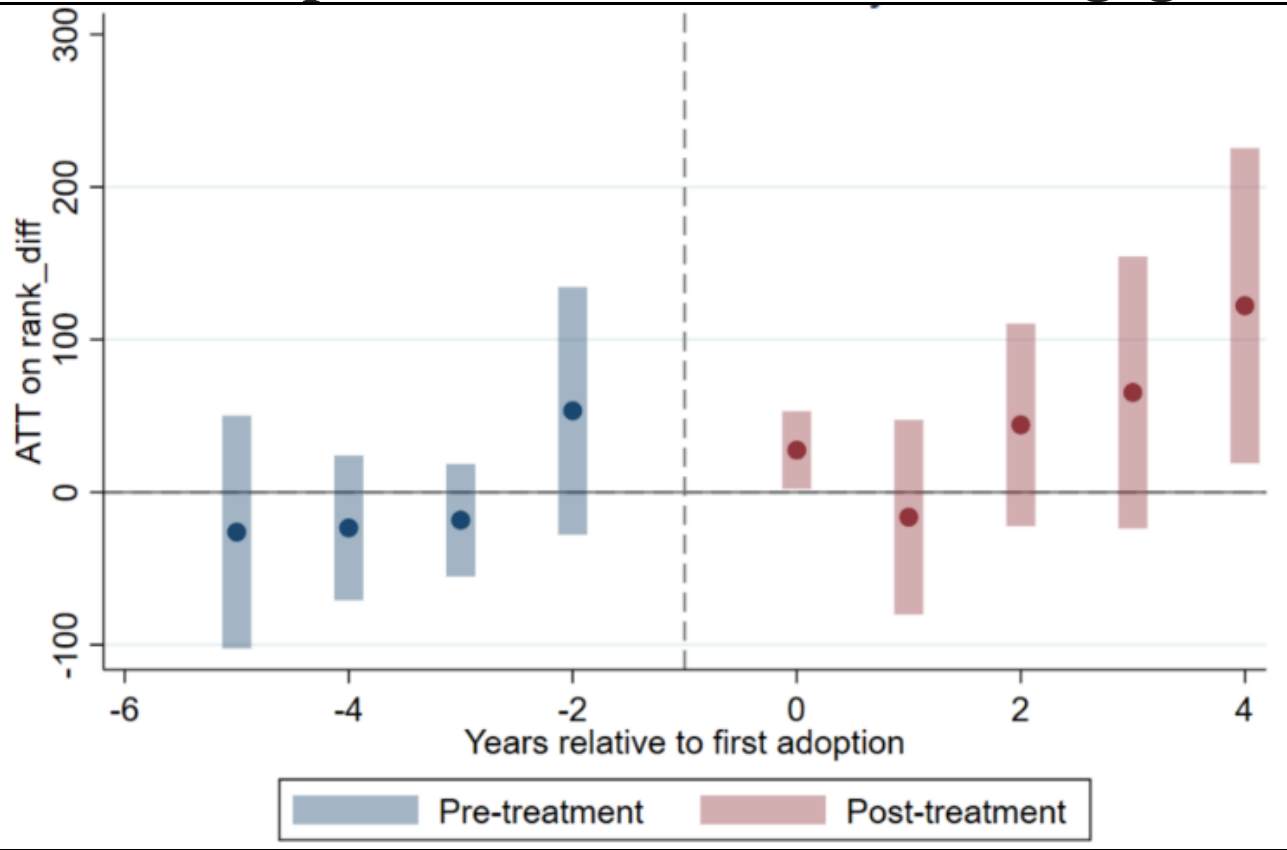

Notes: This figure reports event-study estimates from the Callaway and Sant'Anna staggered difference-in-differences estimator, using not-yet-treated cities as the comparison group. Panel A reports estimates for the full sample. Panels B and C report estimates separately for cities with above- and below-median levels of civic engagement, respectively. The horizontal axis indicates event time relative to the year of official WeChat adoption, and the vertical axis reports the estimated treatment effect on GDP manipulation. Error bars represent 95% confidence intervals. The omitted reference period is the year immediately preceding WeChat adoption.

**Figure S3. Effects of Social Media Disclosure across Civic Engagement Quartiles**

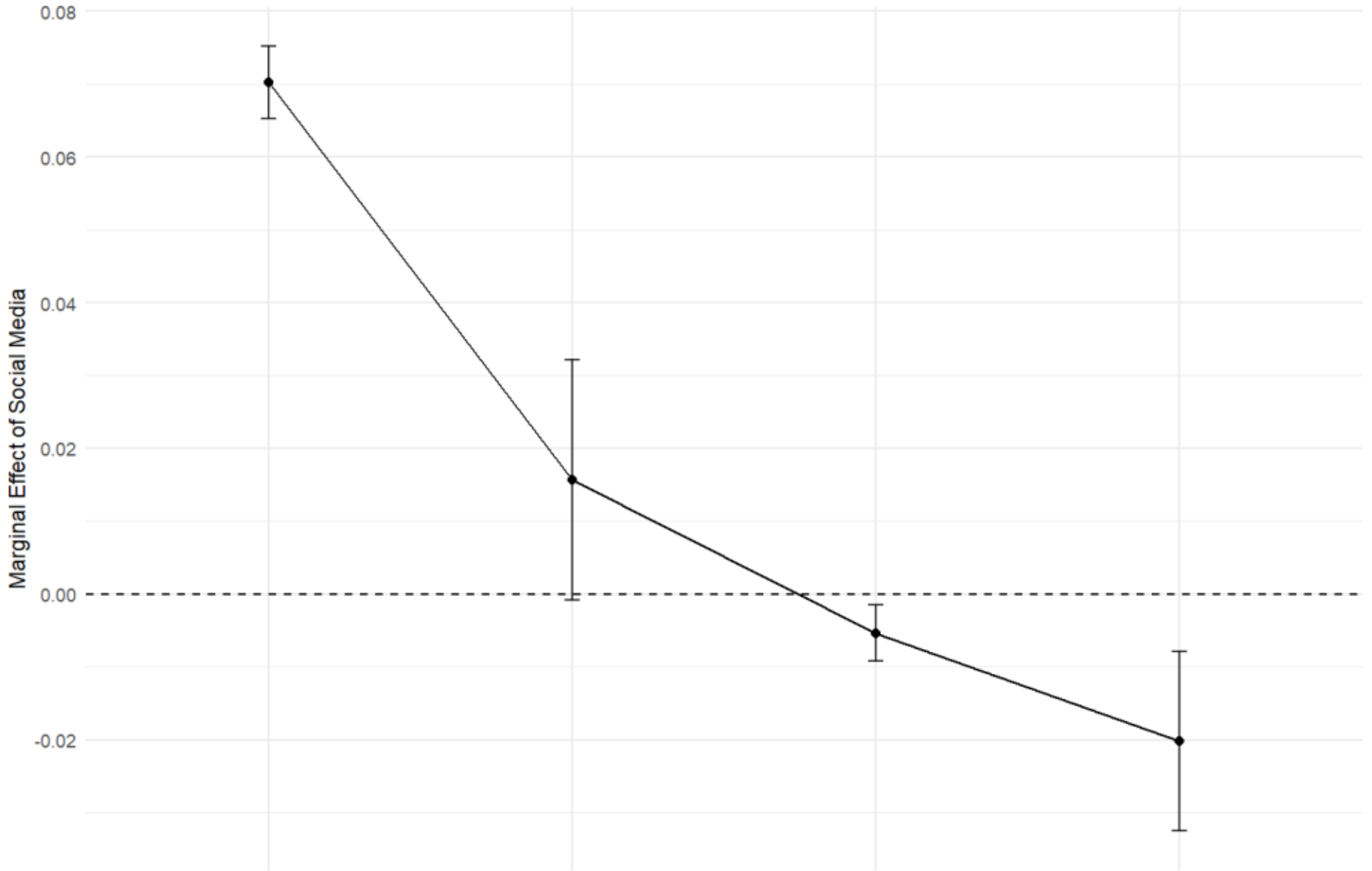

Notes: This figure reports the estimated effects of social media disclosure on GDP manipulation separately across quartiles of the Civic Engagement Index. Observations are divided into four groups based on the 25th, 50th, and 75th percentiles of the Civic Engagement Index, with Quartile 1 representing the lowest level of civic engagement and Quartile 4 the highest. The specifications include the same control variables and fixed effects as the baseline model. Error bars represent 95% confidence intervals.

| Table S1: Keywords for Identifying WeChat Posts | | |
|---|---|---|
| **Topic** | **Keywords (Translation in English)** | **Phrases in Chinese** |
| GDP-related Keywords | GDP, Economy, Investment, Construction, Public Resources, Housing Fund, Consumption, Funds, Income, Finance, Loan, Insurance, Taxation, Subsidy, Price, Enterprise, Salary, Expenditure, Welfare, Wealth Creation, Tax Refund | GDP, 经济, 投资, 建设, 公共资源, 公积金, 消费, 资金, 收入, 财政, 贷款, 保险, 税收, 补贴, 价格, 企业, 工资, 支出, 福利, 致富, 退税 |
| Pollution Emission-related Keywords | Exhaust Gas, Emission, Pollution, Ecology, Environment, Air, Atmosphere, Environmental Protection, Tail Gas, Emission Reduction | 废气, 排放, 污染, 生态, 环境, 空气, 大气, 环保, 尾气, 减排 |

**Table S2: Summary Statistics**

| VARIABLE | Obs. | Mean | Std. Dev. | Min | Max |
|---|---|---|---|---|---|
| **Dependent Variable** | | | | | |
| GDP Manipulation Level | 2,524 | 0.710 | 0.0667 | 0.370 | 1.015 |
| **Independent Variables** | | | | | |
| WeChat Adoption | 2,524 | 0.440 | 0.497 | 0 | 1 |
| log Number of GDP-related Articles | 2,524 | 1.646 | 2.201 | 0 | 6.554 |
| log Number of GDP-related Articles (with Data) | 2,524 | 1.395 | 1.907 | 0 | 6.082 |
| log Number of GDP-related Articles (without Data) | 2,524 | 1.343 | 1.857 | 0 | 6.080 |
| Civic Engagement Index | 2,524 | 0 | 1 | -9.643 | 5.244 |
| **Control Variables** | | | | | |
| Population (in 100,000) | 2,524 | 42.67 | 25.71 | 1.9 | 150 |
| Population Density | 2,524 | 0.0424 | 0.0329 | 0.000 | 0.276 |
| GDP Per Capita | 2,524 | 4.907 | 4.445 | 0.572 | 43.734 |
| Secondary Industry Proportion | 2,524 | 0.469 | 0.107 | 0.106 | 0.893 |
| Tertiary Industry proportion | 2,524 | 0.414 | 0.102 | 0.102 | 0.792 |
| Retail Consumption (in $10^8$ Yuan) | 2,524 | 9,009 | 11,100 | 0.238 | 95,500 |
| Leadership Change | 2,524 | 0.464 | 0.499 | 0 | 1 |
| Mayor Tenure (Years) | 2,524 | 1.756 | 1.122 | 0 | 10 |
| Municipal Party Secretary Tenure (Years) | 2,524 | 2.054 | 1.302 | 0 | 9 |
| **Variables for Placebo Test** | | | | | |
| SO2 Manipulation Level | 2,349 | 0.948 | 2.133 | 0.0848 | 52.557 |
| log Number of Emission-related Articles | 2,349 | 1.235 | 2.000 | 0 | 7.372 |

| Table S3. Decomposition of Variation in the Civic Engagement Index | | | | |
|---|---|---|---|---|
| **Source of Variation** | **Sum of Squares** | **df** | **Mean Square** | **Share of Total Variation** |
| Between cities | 2,036.285 | 291 | 6.998 | 80.7% |
| Within cities | 486.715 | 2,232 | 0.218 | 19.3% |
| **Total** | **2,523** | **2,523** | | **100.0%** |
| F-statistic for equality of city means: 32.09, p-value: < 0.001 | | | | |

Notes: This table decomposes the variation in the Civic Engagement Index into between-city and within-city components using a one-way analysis of variance with city as the grouping variable. The share of total variation is calculated as the corresponding sum of squares divided by the total sum of squares. The F-test evaluates the null hypothesis that mean civic engagement is equal across cities. Between-city differences account for 80.7% of the total variation in the Civic Engagement Index, while within-city changes account for the remaining 19.3%.

| Table S4. Results Using the Alternative Measure of GDP Manipulation | | | | | | | | |
|---|---|---|---|---|---|---|---|---|
| | (1) | (2) | (3) | (4) | (5) | (6) | (7) | (8) |
| DV: GDP Ranking Difference | FE | FE+2SLS | FE | FE+2SLS | FE | FE+2SLS | FE | FE+2SLS |
| WeChat Adoption | -2.705 | **-13.803*** ** | | | | | | |
| | (2.239) | **(5.149)** | | | | | | |
| Log Number of GDP-related Articles | | | -1.016* | **-4.271*** ** | | | | |
| | | | (0.609) | **(1.553)** | | | | |
| Log Number of GDP-related Articles (with Data) | | | | | -1.203* | **-5.541*** ** | -2.955* | **-16.197*** ** |
| | | | | | (0.693) | **(1.694)** | (1.622) | **(5.670)** |
| Log Number of GDP-related Articles (without Data) | | | | | | | 1.884 | **12.139*** |
| | | | | | | | (1.634) | **(6.298)** |
| Control Variables | City FE, Year FE, Population, Population Density, GDP Per Capita, Secondary Industry Proportion, Tertiary Industry Proportion, Retail Consumption, Leadership Change, Mayor Tenure, Municipal Party Secretary Tenure | | | | | | | |
| Observations | 2,524 | 2,524 | 2,524 | 2,524 | 2,524 | 2,524 | 2,524 | 2,524 |
| R-squared | 0.0634 | | 0.0649 | | 0.0651 | | 0.0657 | |

Notes: 1. All columns report the OLS results with city-fixed effect and year-fixed effect. The dependent variable is the GDP ranking difference (the GDP rank using the nighttime light intensity data minus the GDP rank using the government-reported data), which also measures to what extent governments manipulate GDP in reported data. WeChat Adoption is a dummy variable that indicates whether the city has adopted WeChat to publish GDP-related information in the year. Log Number of GDP-related Articles is the log number of GDP-related articles published by each government each year.
2. Columns (2), (4), (6), and (8) use instrumental variable 2SLS. Instrumental variables are the average social media activity of other governments in the same province and year.
3. Clustered standard errors (at the city level) in parentheses: *** p<0.01, ** p<0.05, * p<0.1.

## Table S5. Main Results Excluding Observations from 2017

| | (1) | (2) | (3) | (4) | (5) | (6) | (7) | (8) |
|---|---|---|---|---|---|---|---|---|
| DV: GDP Manipulation | FE | FE+2SLS | FE | FE+2SLS | FE | FE+2SLS | FE | FE+2SLS |
| WeChat Adoption | -0.00579** | **-0.0332*** ** | | | | | | |
| | (0.00255) | **(0.00787)** | | | | | | |
| Log Number of GDP-related Articles | | | -0.00166*** | **-0.00571*** ** | | | | |
| | | | (0.000611) | **(0.00168)** | | | | |
| Log Number of GDP-related Articles (with Data) | | | | | -0.00197*** | **-0.00805*** ** | -0.00592*** | **-0.0367*** ** |
| | | | | | (0.000714) | **(0.00187)** | (0.00177) | **(0.00677)** |
| Log Number of GDP-related Articles (without Data) | | | | | | | 0.00427*** | **0.0332*** ** |
| | | | | | | | (0.00162) | **(0.00725)** |
| Other Controls | City FE, Year FE, Population, Population Density, GDP Per Capita, Secondary Industry Proportion, Tertiary Industry Proportion, Retail Consumption, Leadership Change, Mayor Tenure, Municipal Party Secretary Tenure | | | | | | | |
| Observations | 2,241 | 2,241 | 2,241 | 2,241 | 2,241 | 2,241 | 2,241 | 2,241 |
| R-squared | 0.383 | | 0.385 | | 0.385 | | 0.387 | |

Notes: 1. All columns report the OLS results with city-fixed effect and year-fixed effect. The dependent variable is the GDP manipulation level. WeChat Adoption is a dummy variable that indicates whether the city has adopted WeChat to publish GDP-related information in the year. Log Number of GDP-related articles is the log number of GDP-related articles published by each government each year. The GDP-related articles are further classified into articles that contain specific economic data and articles that do not contain such data.
2. Columns (2), (4), (6), and (8) use instrumental variable 2SLS. Instrumental variables are the average social media activity of other governments in the same province and year.
3. Clustered standard errors (at the city level) in parentheses: *** p<0.01, ** p<0.05, * p<0.1.

**Table S6. Civic Engagement Moderation Results Excluding Observations from 2017**

| | (1) | (2) |
|---|---|---|
| DV: GDP Manipulation | FE | FE |
| Civic Engagement | 0.00903*** | 0.00583*** |
| | (0.00197) | (0.00182) |
| WeChat Adoption | -0.00563** | |
| | (0.00256) | |
| Civic Engagement * WeChat Adoption | -0.00613*** | |
| | (0.00146) | |
| Log Number of GDP-related Articles | | -0.00313** |
| | | (0.00129) |
| Civic Engagement * Log Number of GDP-related Articles | | -0.00221*** |
| | | (0.000543) |
| Other Controls | City FE, Year FE, Population, Population Density, GDP Per Capita, Secondary Industry Proportion, Tertiary Industry Proportion, Retail Consumption, Leadership Change, Mayor Tenure, Municipal Party Secretary Tenure | |
| Observations | 2,241 | 2,241 |
| R-squared | 0.392 | 0.390 |

Notes:1. All columns report the OLS results with city-fixed effect and year-fixed effect. The dependent variable is the GDP manipulation level. Civic Engagement is standardized. WeChat Adoption is a dummy variable that indicates whether the city has adopted WeChat to publish GDP-related information in the year. Log Number of GDP-related Articles is the log number of GDP-related articles published by each government each year.

2. Clustered standard errors (at the city level) in parentheses: *** p<0.01, ** p<0.05, * p<0.1

**Table S7. Pollution-Emission Placebo Test Excluding Observations from 2017**

| | (1) | (2) | (3) | (4) | (5) | (6) | (7) | (8) |
|---|---|---|---|---|---|---|---|---|
| DV: $SO_2$ Manipulation | FE | FE+2SLS | FE | FE+2SLS | FE | FE+2SLS | FE | FE+2SLS |
| WeChat Adoption | -0.273 | **-0.0907** | | | | | | |
| | (0.260) | **(0.0727)** | | | | | | |
| Log Number of Emission-related Articles | | | -0.0477 | **-0.201** | | | | |
| | | | (0.0439) | **(0.201)** | | | | |
| Log Number of GDP-related Articles | | | | | -0.0315 | **0.0100** | | |
| | | | | | (0.0280) | **(0.0283)** | | |
| Log Number of GDP-related Articles (with Data) | | | | | | | -0.0330 | **0.0205** |
| | | | | | | | (0.0293) | **(0.0364)** |
| Control Variables | City FE, Year FE, Population, Population Density, GDP Per Capita, Secondary Industry Proportion, Tertiary Industry Proportion, Retail Consumption, Leadership Change, Mayor Tenure, Municipal Party Secretary Tenure | | | | | | | |
| Observations | 2,102 | 2,102 | 2,102 | 2,102 | 2,102 | 2,102 | 2,102 | 2,102 |
| R-squared | 0.0359 | | 0.0345 | | 0.0338 | | 0.0337 | |

Notes: 1. All columns report the OLS results with city-fixed effect and year-fixed effect. The dependent variable is the $SO_2$ emission manipulation level. WeChat Adoption is a dummy variable that indicates whether the city has adopted WeChat to publish articles in the year. Log Number of Emission-related Articles is the log number of Emission-related articles published by each government each year. The sample size has slightly decreased due to the absence of $SO_2$ data released by some city governments.

2. Columns (2), (4), (6), and (8) report 2SLS estimates using the corresponding provincial averages as instruments for WeChat adoption, emissions-related posts, GDP-related posts, and GDP-related posts containing economic data, respectively.

3. Clustered standard errors (at the city level) in parentheses: *** $p<0.01$, ** $p<0.05$, * $p<0.1$

## Table S8. Alternative GDP Manipulation Measure Excluding Observations from 2017

| | (1) | (2) | (3) | (4) | (5) | (6) | (7) | (8) |
|---|---|---|---|---|---|---|---|---|
| DV: GDP Ranking Difference | FE | FE+2SLS | FE | FE+2SLS | FE | FE+2SLS | FE | FE+2SLS |
| WeChat Adoption | -3.665* | **-16.366*** ** | | | | | | |
| | (2.029) | **(4.753)** | | | | | | |
| Log Number of GDP-related Articles | | | -1.170** | **-4.894*** ** | | | | |
| | | | (0.541) | **(1.454)** | | | | |
| Log Number of GDP-related Articles (with Data) | | | | | -1.303** | **-5.856*** ** | -2.050 | **-12.042** ** |
| | | | | | (0.615) | **(1.581)** | (1.261) | **(4.776)** |
| Log Number of GDP-related Articles (without Data) | | | | | | | 0.808 | **7.167** |
| | | | | | | | (1.270) | **(5.352)** |
| Control Variables | City FE, Year FE, Population, Population Density, GDP Per Capita, Secondary Industry Proportion, Tertiary Industry Proportion, Retail Consumption, Leadership Change, Mayor Tenure, Municipal Party Secretary Tenure | | | | | | | |
| Observations | 2,241 | 2,241 | 2,241 | 2,241 | 2,241 | 2,241 | 2,241 | 2,241 |
| R-squared | 0.0632 | | 0.0653 | | 0.0650 | | 0.0652 | |

Notes: 1. All columns report the OLS results with city-fixed effect and year-fixed effect. The dependent variable is the GDP ranking difference (the GDP rank using the nigh-light intensity data minus the GDP rank using the government-reported data), which also measures to what extent governments manipulate GDP in reported data. WeChat Adoption is a dummy variable that indicates whether the city has adopted WeChat to publish GDP-related information in the year. Log Number of GDP-related Articles is the log number of GDP-related articles published by each government each year.
2. Columns (2), (4), (6), and (8) use instrumental variable 2SLS. Instrumental variables are the average social media activity of other governments in the same province and year.
3. Clustered standard errors (at the city level) in parentheses: *** p<0.01, ** p<0.05, * p<0.1.